\documentclass[tightenlines,floatfix,superscriptaddress,twocolumn,floatfix,pra,10pt]{revtex4-2}
\usepackage{graphicx}
\usepackage{amssymb}
\usepackage{amsmath}
\usepackage{color}
\usepackage{xspace}

\newcommand{\ket}[1]{| #1 \rangle}

\newcommand{\rb}[1]{\left( #1 \right)}
\newcommand{\ew}[1]{\langle #1 \rangle}
\newcommand{\beq}{\begin{eqnarray}}
\newcommand{\eeq}{\end{eqnarray}}
\newcommand{\eq}[1]{Eq.~(\ref{#1})}
\newcommand{\fig}[1]{Fig.~\ref{#1}}
\newcommand{\secref}[1]{Sec.~\ref{#1}}

\newcommand{\etal}{{\em et al.}\xspace}

\begin{document}
\title{Asymmetric arms maximise visibility in hot-electron interferometers}
\author{Clarissa J. Barratt}
\affiliation{ 
Joint Quantum Centre Durham-Newcastle,
School of Mathematics, Statistics and Physics,
Newcastle University,
Newcastle upon Tyne NE1 7RU, UK
}
\author{Sungguen Ryu}
\affiliation{ 
Institute for Cross-Disciplinary Physics and Complex Systems IFISC (UIB-CSIC), E-07122 Palma de Mallorca, Spain
}
\author{Lewis A. Clark}
\affiliation{ 
Joint Quantum Centre Durham-Newcastle, 
School of Mathematics, Statistics and Physics,
Newcastle University,
Newcastle upon Tyne NE1 7RU, UK
}
\affiliation{
Centre for Quantum Optical Technologies, 
Centre of New Technologies, 
University of Warsaw, 
Banacha 2c, 
02-097 Warsaw, Poland
}
\author{H.-S. Sim}
\affiliation{Department of Physics, Korea Advanced Institute of Science and Technology, Daejeon 34141, Republic of Korea}
\author{Masaya Kataoka}
\affiliation{National Physical Laboratory, Hampton Road, Teddington, Middlesex TW11 0LW, United Kingdom}
\author{Clive Emary}
\affiliation{ 
Joint Quantum Centre Durham-Newcastle,
School of Mathematics, Statistics and Physics,
Newcastle University,
Newcastle upon Tyne NE1 7RU, UK
}
\date{\today}
\begin{abstract}
  We consider theoretically an electronic Mach-Zehnder interferometer constructed from quantum Hall edge channels and quantum point contacts, fed with single electrons from a dynamic quantum dot source.
  By considering the energy dependence of the edge-channel guide centres, we give an account of the phase averaging in this set up that is particularly relevant for the short, high-energy wavepackets injected by this type of electron source.
  We present both analytic and numerical results for the energy-dependent arrival time distributions of the electrons and also give an analysis of the delay times associated with the quantum point contacts and their effects on the interference patterns.
  A key finding is that, contrary to expectation, maximum visibility requires the interferometer arms to be different in length, with an offset of up to a micron for typical parameters. By designing interferometers that incorporate this asymmetry in their geometry, phase-averaging effects can be overcome such that visibility is only limited by other incoherent mechanisms.
\end{abstract}

\maketitle

\section{Introduction}

The ability to perform interferometry experiments with electrons opens up new pathways to investigate the fundamental mechanisms of relaxation and decoherence in the solid state \cite{Grenier2011,Bocquillon2014}.
The first realization of an electronic Mach-Zehnder interferometer (MZI) \cite{Ji2003} set the template for future electron quantum optics experiments \cite{Neder2006,Litvin2007,Roulleau2008a,Roulleau2008b,Roulleau2009,Bieri2009,Tewari2016} with quantum Hall edge channels acting as electron waveguides and quantum point contacts taking the place of beamsplitters.
In these experiments, electrons were injected from DC sources at energies close to the Fermi level.  The decoherence mechanisms of such electrons have been discussed extensively \cite{Marquardt2004, Chung2005, Neuenhahn2008} with the Coulomb interaction through plasmon emission \cite{DasSarma1982, Gasser1987,Hu1993, Giuliani2005} thought to be the dominant mechanism \cite{Neder2008,Youn2008, Levkivskyi,Degiovanni2009}.
 
Mirroring the advent of single-photon sources in quantum optics \cite{Eisaman2011,Senellart2017}, sources of single electrons have also been developed.  Driven mesoscopic capacitors  \cite{Gabelli2006,Feve2007} have been used to successfully realise single-electron quantum-optics experiments \cite{Bocquillon2014} such as Hanbury Brown-Twiss \cite{Bocquillon2012}, and Hong-Ou-Mandel \cite{Bocquillon2013,Freulon2015}.
In this work we consider a different type of single-electron source: the gate-modulated, or dynamic, quantum dot \cite{Fujiwara2008, Kaestner2008, Kaestner2015}, originally developed for metrology purposes \cite{Giblin2012}.
In contrast with DC sources or mesoscopic capacitors, dynamic quantum dots inject electrons at energies significantly in excess of the Fermi level.  The spatial separation that this produces between injected ``hot'' electrons and the bulk is thought to suppress the Coulomb interactions \cite{Clark2020}, and this opens up the possibility that different mechanisms are important for the loss of coherence of hot electrons than for ``cold'' ones. 
The emission of longitudinal-optical (LO) phonons is certainly an important relaxation process \cite{Fletcher2013, Emary2016} and emission of acoustic phonons is also believed to play a role \cite{Johnson2018,Emary2019,Ota2019}.  In Ref.~[\onlinecite{Clark2020}], these same phonon processes were also predicted to be dominant in causing decoherence of hot electrons in an interferometer, although it was shown that the strengths of these effects could be minimised by correct parameter choice and deployment of filtration schemes.

In this paper we focus on the issue of phase averaging in hot-electron MZIs.  Whilst this topic has been discussed for cold electrons, both from DC \cite{Chung2005} and driven-capacitors sources \cite{Haack2011}, our analysis here is specifically relevant for temporally-short, high-energy single-electron wavepackets.
The short duration of these wavepackets makes matching their arrival at the second MZI beamsplitter essential for observation of interference. Thus successful interferometer design relies on a detailed understanding of the arrival times in this system.
The central finding of this work is that, contrary to expectations from the optical MZI as well as from Refs~\cite{Chung2005,Haack2011,Beggi2015}, for the arrival times to match and thus interferometric visibility be maximised,  the two arms of the MZI should be of different lengths. Indeed, for typical parameters, this length offset could be up to 1\,$\mu$m and thus significant for interferometer design.
The origin of this effect is a combination of two factors. Firstly, the high energy of the electrons imparts them with a high wave number and thus an enhanced sensitivity to changes in path lengths.
Secondly, for electrons in the quantum Hall regime, the position of the electron guide centre (equivalent to the position of optical path) is dependent on the energy of the electron, with the result that different energy components in an electron wavepacket travel slightly different paths in the MZI and thus pick up different phases.  This additional energy-dependence of the phases results in a difference in the travel times of the electrons around the two arms of the interferometer, and necessitates an offset in the path lengths to compensate.

In addition to this effect, we also give an analysis of how the properties of quantum point contact beamsplitters can affect the arrival times of the electrons and show that only in the case where the beamsplitters are energetically narrow and significantly asymmetric do they become important in determining the interference patterns.

The  paper proceeds as follows.  In \secref{sec:model} we introduce our model of the MZI and the scattering approach we use to describe its properties.  
Section~\ref{SEC:MZIphases} describes the phases accumulated by electrons traversing the MZI, and in \secref{sec:delayTimesandVisibility} we derive analytic expressions for the energy-dependent arrival time distribution of electrons. From this features such as electron travel times and the visibility of interference fringes are determined.
In \secref{sec:numericalResults}, we present numerical results for the arrival time distributions across a greater range of parameters, and in \secref{SEC:roleBS} we consider the role that delays at the beamsplitters play in determining interference in the MZI. 
We conclude with discussions in \secref{sec:discussion}. 
Details of our parameter choices are given in Appendix~\ref{SEC:params} and the scattering properties of the beamsplitters are calculated in Appendix~\ref{SEC:BSphase}.

\section{Model \label{sec:model}}

\begin{figure}[tb]
\centering
 \includegraphics[width=\columnwidth]{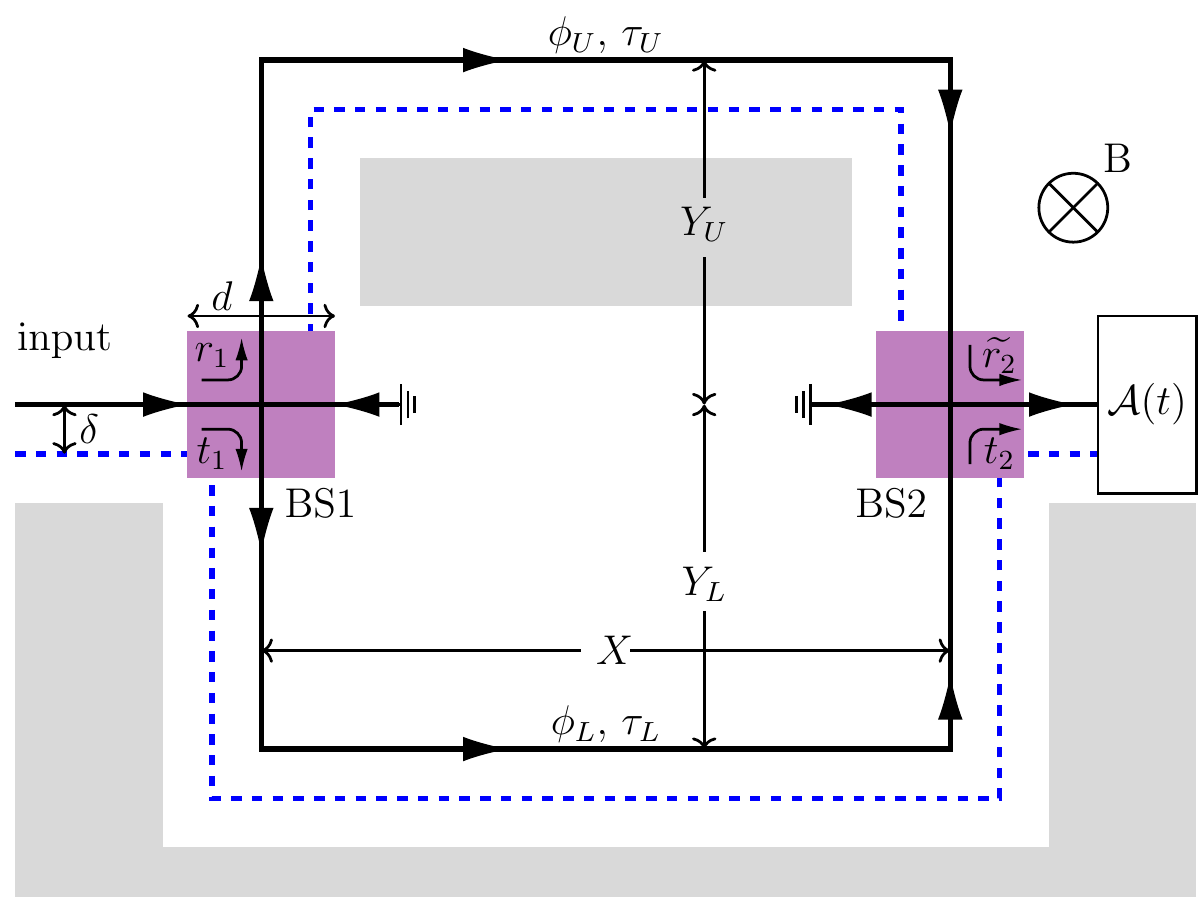}
  \caption{
    Schematic of an electronic Mach-Zehnder interferometer. Grey represents the relevant areas of depletion of the 2DEG and purple regions represent beamsplitters realised by quantum point contacts. 
    Electrons are injected on the left and collected on the right, where arrival time distribution $\mathcal{A}(t)$ is measured.
    The solid-black and dashed-blue lines represent the paths taken by electrons with wave number $k_0$ and $k_0+\Delta k$ respectively. 
    For $\Delta k>0$ the path of the electron moves closer to the sample edge by a distance $\delta >0$, and this changes the phases $\phi_U$ and $\phi_L$ accumulated by electrons on the upper and lower arms. The energy dependence of the phases gives rise to the corresponding travel times of electron wavepackets $\tau_{U/L}$.
    Also indicated are the lengths $X$, $Y_U$ and $Y_L$ of parts of the interferometer arms, and $d$ the size of the beamsplitter (BS) regions.
  }
  \label{fig:mzi}
\end{figure}

We consider an electronic MZI with wave guides realised as described in \fig{fig:mzi}.
We take the transverse confinement in the edge channels to be harmonic \cite{Datta1997,Emary2016} with confinement frequency $\omega_y$.  Assuming that our electrons are always confined to the lowest Landau level, their dispersion relative to the subband bottom is
\begin{equation}
    E_{k} = 
    \rb{\frac{\omega_y}{\Omega}}^2 
    \frac{\hbar^2 k^2}{2m_e^*  } 
    \label{EQ:energy_first}
    ,
\end{equation} 
where we have defined
$
  \Omega^2 = \omega_y^2 + \omega_c^2
$
with the cyclotron frequency
$
 \omega_c = eB/m_e^*
$
in terms of the effective mass $m_e^*$ and charge $e>0$ of the electron, as well as the magnetic field strength $B>0$.
  
We consider electrons to be injected in a coherent wavepacket with momenta distributed according to a Gaussian profile \cite{ryu:ultrafast}
with central wave number $k_0$ and with width parameter $\alpha$.  Just prior to injection into the MZI, the electronic state is thus
\beq
  \ket{\Psi(0)} 
   = \rb{\frac{\alpha L^2}{2\pi^3}}^{1/4} \int dk \,e^{-\alpha(k-k_0)^2} \ket{0,k}
  \label{EQ:startstate}
  ,
\eeq
where $\ket{0,k}$ is the lowest Landau level eigenstate of wave number $k$.  In a position representation with $x$ the transport direction and $y$ transverse to it with $y=0$ defined as the potential minimum, these eigenfunctions read
$
  \psi_{0,k}(x,y) = 
  \ew{x,y|0k} =
  \frac{1}{\sqrt{L}} e^{ikx}\chi_{0,k}\rb{y} 
$
where $L$ is a quantization length and 
$
  \chi_{0k}\rb{y} 
  =
  \rb{\pi l_\Omega^2}^{-1/4}
  e^{
    - \frac{1}{2}\left[y-y_G(k)\right]^2/l_\Omega^2
  }
$ is the transverse wave function, which is expressed in terms of the confinement length
$
  l_\Omega = \sqrt{\hbar/ \rb{m_e^* \Omega}}
$, and the guide centre of the electron trajectory
\beq
  y_\text{G}(k) =  -\frac{\omega_c^2}{\Omega^2}\frac{\hbar {k} }{eB} 
  \label{EQ:guidecentre}
  .
\eeq
Here we have worked in the Landau gauge such that the vector potential reads $\mathbf{A}=\rb{By,0,0}$.  By considering the probability density associated with \eq{EQ:startstate}, we find that the parameter $\alpha$ can be expressed in terms of the energetic width $\sigma_E$ of the initial wavepacket as
$
  \alpha = \rb{\hbar v_0 /2\sigma_E}^2
$ such that the localization length of the wave packet in the transport direction is $\sigma_l = \hbar v_0 /\sigma_E $.

\subsection{Scattering approach to MZI}

We take into account the effect of the MZI on this wavepacket within a scattering approach \cite{Chung2005,Haack2011,Beggi2015}.  The action of the two beamsplitters $i=1,2$ is described by the scattering matrices
\beq
  S_i  
  =
  \rb{\begin{array}{cc}
    r_i
      & \widetilde{t}_i \\ 
    t_i
      & \widetilde{r}_i
  \end{array}}
  =
  \rb{\begin{array}{cc}
    \sqrt{R_i} e^{i\rho_i}
      & \sqrt{T_i} e^{i\widetilde{\theta}_i} \\ 
    \sqrt{T_i} e^{i \theta_i}
      & \sqrt{R_i} e^{i\widetilde{\rho}_i}
  \end{array}}
  \label{EQ:scat}
  ,
\eeq
where $t_i$, $\widetilde{t}_i$ $r_i$, and $\widetilde{r}_i$ are scattering amplitudes, the sense of which are shown in  \fig{fig:mzi}, and where we have transmission $T_i$ and reflection $R_i$ probabilities obeying $R_i=1-T_i$, and phases obeying the unitarity condition of 
$
  \theta_i + \widetilde{\theta}_i - \rho_i -\widetilde{\rho}_i = (2n+1) \pi;
  ~n \in \mathbb{Z}
$.
We note that this scattering matrix, and all its constituent phases and amplitudes, are functions of $k$ here, but we leave this dependence implicit to avoid over-burdening the notation.
Assuming that electrons travelling the upper and lower arms of the interferometer acquire phases $\phi_U$ and $\phi_L$ respectively, the total scattering amplitude for transmission through the MZI is 
\beq
  t^\mathrm{MZI}_k
  &=&
  \sqrt{R_1 R_2} e^{i\rb{\rho_1 + \widetilde{\rho}_2 + \phi_U}} 
  + 
  \sqrt{T_1T_2} e^{i \rb{\theta_1 + \theta_2 + \phi_L}}
  \label{EQ:scatamp}
  ,
\eeq
where, again, all quantities depend on wave number $k$.
With the initial state (\ref{EQ:startstate}) and scattering amplitude (\ref{EQ:scatamp}), the electron probability distribution at the output of the MZI reads
\beq
  P(x,t)
  &=&
   \sqrt{\frac{1}{2\pi^3}}
  \int dy \, dk\,  dk' \,
   \rb{t^\mathrm{MZI}_{k'}}^*
   t^\mathrm{MZI}_{ k} 
   \chi^*_{0k'}(y) \chi_{0k}(y)
   \nonumber\\
   &&
   \times
   e^{-i (E_k-E_{k'})t/\hbar + i(k-k')x}
   e^{-\alpha[(k-k_0)^2+(k'-k_0)^2]},
   \nonumber
\eeq
where time $t$ is measured from the injection time and position $x$ is measured from the exit.
With a detector position of $x_D$ and assuming a narrow wavepacket such that the velocity is assumed constant over the wavepacket, the arrival time distribution is approximately $\mathcal{A}(t) = v_0 P(x_D, t)$ \cite{Emary2016}, where 
\beq
  v_0 = \frac{1}{\hbar} \left. \frac{d E_k}{dk} \right|_{k=k_0} 
  = \frac{ \hbar \omega_y^2 k_0 }{m_e^* \Omega^2}
  ,
\eeq
is the velocity of the wavepacket \cite{Kataoka2016}.

Thus we obtain
\beq
 \mathcal{A}(t) &=&  v_0 
 \int dk\,  dk' \,
   \rb{t^\mathrm{MZI}_{k'}}^*
   t^\mathrm{MZI}_{ k} 
   e^{- \frac{1}{4}\mu^2
                (k-k')^2} \nonumber\\
   &&
   \times
   e^{-i (E_k-E_{k'})t/\hbar + i(k-k')x_D}
   e^{-\alpha[(k-k_0)^2+(k'-k_0)^2]},
   \nonumber\\
   \label{EQ:Pxt2}
\eeq
with $ \mu =  l_\Omega \omega_c/ \Omega$.

\section{MZI phases \label{SEC:MZIphases}}

Central to the analysis here is the energy dependence of the phases $\phi_U$ and $\phi_L$ of \eq{EQ:scatamp} picked up by the electron along the interferometer arms.  Here we adopt the rectilinear geometry of \fig{fig:mzi} and take the arms of the interferometer to exclude the beamsplitter regions.  Figure~\ref{fig:mzi} defines the lengths $X$, $Y_U$, $Y_L$ and $d$ of different parts of the interferometer.
We assume that electrons with wave number $k_0$ at the centre of the wavepacket are partitioned by the beamsplitters into both channels and that the paths followed by these electrons are given by the solid black lines in the figure.
For electrons with wave number $k=k_0 +\Delta k$, we see from \eq{EQ:energy_first} that a change in wave number results in a shifted guide centre such that the electron follows a slightly different path around the MZI.  This altered path, indicated in the figure by the blue dashed line, is displaced relative to the $k_0$ path a distance of 
$
  \delta  = 
   \rb{\frac{\omega_c}{\Omega}}^2
   \frac{\hbar}{\rb{eB}} \Delta k 
$ which we define being positive for displacements towards the edge for $\Delta k>0$.
Along these paths, the electron accumulates both the dynamical and magnetic phases.
The dynamical phases are given by the product of the wave number with the path length, but here the above change in guide centre means that the path length is also dependent on $k$. For the geometry of \fig{fig:mzi}, we therefore obtain the dynamical phases
\beq
  \phi_p^{(\mathrm{dyn})}
  =
  (k_0 + \Delta k)(2 Y_p - d + X -4\xi_p  \delta )
  \label{eq:dynphases}
  ,
\eeq
where $p\in \left\{U,L\right\}$ and where $\xi_p$ is a sign factor that takes the value $\xi_U=+1$ on the upper path and $\xi_L=-1$ on the lower.
Meanwhile the magnetic phase is proportional to the line integral of the vector potential along the paths in question.  With Landau gauge as before, we find the  phases accumulated along the two arms to be
\footnote{In calculating these magnetic phases, consistency with the wave functions introduced earlier requires a coordinate offset $y_0$ between the central electron path and the potential minimum in the horizontal channels.  However, as this essentially corresponds to a redefinition of the gauge, it cancels in the phase difference between upper and lower channels and has no observable consequence. We therefore set $y_0=0$ for simplicity.
}
\beq
  \phi_p^{(\mathrm{mag})}
  =
  -\frac{eB}{\hbar} 
  \rb{\xi_p Y_p - \delta}
  \rb{ X - 2\xi_p \delta}
  \label{eq:ABphases}
  .
\eeq
Note that the motion within the beamsplitter regions is explicitly excluded from the calculation of these phases. All this has assumed is that the entry and exit points are as in \fig{fig:mzi}.

\section{Travel times and visibility \label{sec:delayTimesandVisibility}}
Analytic progress can then be made by recalling that we have a narrow wavepacket, $\alpha^{-1/2} \ll k_0$, such that $\Delta k \ll k_0$ for all states in the wavepacket.  A further consequence of this is that $|\delta|$ will be  similarly small compared with the dimensions of the interferometer (formally we have  
$
|\delta| \sim  l_c^2/\sqrt{\alpha} \ll (2Y_{U/L}+X)
 $ with $l_c=\sqrt{\hbar/m_e^* \omega_c}$ the cyclotron length
).
This then means that the term proportional to $(\Delta k)\delta$ in \eq{eq:dynphases} can be neglected.
To approximate the magnetic phases, we consider the difference between them --- the gauge-independent Aharonov-Bohm phase \cite{Aharonov1959} --- which here reads
\beq
  \phi_U^{(\mathrm{mag})}-\phi_L^{(\mathrm{mag})}
  =
  -
  \frac{eB}{\hbar} 
  \left\{
    a_0 
    - \delta \,l_0 + 4 \delta^2
  \right\}
  ,
\eeq
in terms of the central path difference,
$
  l_0 = 2(Y_U - Y_L)
$, and area enclosed by the central paths.
$
  a_0 = X(Y_U+Y_L)
$.  The first term clearly recovers the expected Aharonov-Bohm phase of the MZI loop described in \fig{fig:mzi}.
The quadratic term will be negligible when $|\delta| \ll l_0$. As we will see shortly, the operating point that yields the maximum visibility for the interferometer is obtained when $l_0 \sim 1\mu$m, such that for the region of interest we have $|\delta| \ll l_0$ and we therefore neglect the term proportional to $\delta^2$ in \eq{eq:ABphases}.

Summing \eq{eq:dynphases} and \eq{eq:ABphases} and neglecting these terms then, we obtain total phases picked up along the arms 
\beq
  \phi_p &\approx &
    -\xi_p \frac{eB}{\hbar} X Y_p +
    k_0(2 Y_p - d + X) 
    \nonumber\\
    &&
    + 
    \Delta k
    \left\{ 
    \frac{\Omega^2 + \omega_c^2}{\Omega^2}
      \rb{2 Y_p + X} 
      -d -
       4 \xi_p  \frac{k_0 l_\Omega^2 \omega_c}{\Omega}
    \right\}
    .
  \nonumber
\eeq
We then also assume that, compared with the phases, the beamsplitter transmission probabilities are slowly-varying functions of $k$ and thus we can evaluate them at the central wave number $k_0$; $T_i \to T_{i}^{(k_0)}$ and $R_i \to R_{i}^{(k_0)}$. Then, linearising the dispersion, $E_k \simeq E_{k_0} + \hbar v_0 \Delta k$ and integrating, \eq{EQ:Pxt2} evaluates as
\begin{widetext}
\beq
  \mathcal{A}(t)
  &\approx&
  \sqrt{\frac{v_0^2}{\pi \rb{2\alpha + \mu^2} }}
  \left\{
   R_1^{(k_0)} R_2^{(k_0)}
  \exp\left[
    - \frac{\left\{x_D-v_0 (t-\tau_U)\right\}^2}{2\alpha + \mu^2}
  \right]
  +
  T_1^{(k_0)} T_2^{(k_0)}
  \exp\left[
    - \frac{\left\{x_D-v_0 (t-\tau_L)\right\}^2}{2\alpha + \mu^2}
  \right]
  \right.
  \nonumber\\
  &&
  ~~~~~~~~~~~~
  \left.
  +
  2
  \sqrt{R_1^{(k_0)} R_2^{(k_0)} T_1^{(k_0)} T_2^{(k_0)}}
  \exp\left[
    - 
    \frac{v_0^2\rb{\tau_U-\tau_L}^2}{8\alpha}
    - 
    \frac{\left\{x_D-v_0 \rb{t - \frac{1}{2}\tau_L-\frac{1}{2}\tau_U}\right\}^2}{2\alpha + \mu^2}
  \right]
   \cos\Phi_0
  \right\}
  \label{EQ:PUxt}
  .
\eeq
\end{widetext}
Thus, the arrival time distribution shows two moving lobes, resulting from travel round the upper and lower paths, and an interference term between them.  The travel time of the two lobes is given by
\beq
  \tau_p
  &=& 
  \left. \hbar \frac{d \Phi_p}{dE}\right|_{E=E0} 
  = 
  \left. \frac{1}{v_0} \frac{d\Phi_p}{dk} \right|_{k=k_0}
  \label{EQ:firstdelaysbydiff}
  ,
\eeq
where
$
  \Phi_U
  = \rho_1+ \widetilde{\rho}_2 + \phi_U
$
and
$
  \Phi_L =
  \theta_1 + \theta_2+ \phi_L 
$.
The interference term shows oscillations as a function of the central phase difference  
\beq
\Phi_0
  =
  \left[ \Phi_U - \Phi_L\right]_{k=k_0}
  =
   -\frac{e B}{\hbar} a_0
  + k_0 l_0 
  + \Phi_0^\mathrm{BS}
  ,
\eeq
where we have separated the beamsplitter contribution
\beq
 \Phi_0^\mathrm{BS} 
  &=&
  \left[
  \rho_1 
  + \widetilde{\rho_2}
  - \theta_1
  - \theta_2
  \right]_{k=k_0}
  .
\eeq

Integrating \eq{EQ:PUxt} over time, we obtain the total probability of detection at the output to be 
\beq
  P_\mathrm{tot}
  &=&
   R_1^{(k_0)} R_2^{(k_0)}
  +
  T_1^{(k_0)} T_2^{(k_0)}
  \nonumber\\
  &&
  +
  2
  \sqrt{R_1^{(k_0)} R_2^{(k_0)} T_1^{(k_0)} T_2^{(k_0)}}\,
   \mathcal{D} \,
   \cos\rb{\Phi_0}
   \label{EQ:Ptot}
\eeq
where
$
  \mathcal{D} = \exp\left[
    - 
    v_0^2(\Delta\tau)^2/ \rb{8\alpha}
  \right]
$, depends on the difference in travel times 
\beq
  \Delta \tau &=& \tau_U-\tau_L
  \nonumber\\
  &=& 
  v_0^{-1} 
  \frac{d}{dk}
  \left[
    \phi_U-  \phi_L
  \right]_{k=k_0}
  +  \Delta \tau_\mathrm{BS}
  .
\eeq
with
\beq
  \Delta \tau_\mathrm{BS} =  v_0^{-1}\frac{d}{dk}\rb{\rho_1+\widetilde{\rho_2} - \theta_1-\theta_2}
  \label{eq:asymm_delays}
  ,
\eeq
the contribution due to the beamsplitters.
The interferometric visibility of oscillations displayed by \eq{EQ:Ptot} is 
\beq
  \mathcal{V} = \mathcal{D}\times \frac{ 2
  \sqrt{R_1^{(k_0)} R_2^{(k_0)} T_1^{(k_0)} T_2^{(k_0)}}}{R_1^{(k_0)} R_2^{(k_0)}
  +
  T_1^{(k_0)} T_2^{(k_0)}} 
  \label{eq:full-expression}
\eeq
such that $ \mathcal{D}$ is observed to be the non-trivial contribution to the visibility that arises from phase averaging (for 50:50 beamsplitter, $\mathcal{V}\to\mathcal{D}$).
Writing 
$
  \alpha = \rb{\hbar v_0 /2\sigma_E}^2
$
in terms of the energetic width $\sigma_E$ of the initial wavepacket, we can rewrite this phase-averaging factor as
\beq
  \mathcal{D}
  &=&
  \exp 
  \left\{
    -\frac{1}{2} 
    \rb{
      \frac{
        l_0 - l_\mathrm{offset}
      }{
      \Sigma_l
      }
    }^2    
  \right\}
  \label{eq:visibility}
\eeq
in terms of the path difference $l_0$, the ``offset length''
\beq
  l_\mathrm{offset} =
  \frac{ \Omega^2  v_0}{\Omega^2+ \omega_c^2}
  \left[
    \frac{8 \omega_c}{\omega_y^2} -  \Delta \tau_\mathrm{BS} 
  \right]
  \label{EQ:loffset}
  ,
\eeq
and an effective single-particle coherence length of the electron wavepacket
\beq
  \Sigma_l
  =
  \frac{\Omega^2}{\Omega^2 + \omega_c^2}
  \frac{\hbar v_0}{\sigma_E }
  \label{EQ:visibilitywidth}
  .
\eeq 
Thus, our theory predicts a modulation of the visibility with a Gaussian function of the path difference $l_0$ with offset $l_\mathrm{offset}$ and characteristic width $\Sigma_l$.
In a basic quantum-optics MZI, with no AB phase and a static path for all particles and energy-independent beamsplitters, we would expect a similar expression for the phase-averaging effects but with $l_\mathrm{offset}=0$, indicating that the maximum visibility occurs when the path difference is zero.  Here, however, maximum visibility requires  
$
    l_0 =  l_\mathrm{offset}
$, indicating that a difference in the length of the two interferometer arms is necessary for optimal coherence. A further difference from these naive considerations is that the width of the visibility feature here is determined by the effective coherence length $\Sigma_l$, which in general is different to the localisation length of the initial wavepacket. Comparison of these quantities yields $\frac{1}{2}\sigma_l \le \Sigma_L \le \sigma_l$ with the righthand limit approached in the large-field limit.

Addressing the beamsplitter contribution in the above, if we make no further assumptions about the beamsplitters other than that they act symmetrically on electrons coming from different directions, the beamsplitter phases obey
\beq
   \rho_i
   = \widetilde{\rho}_i
   = \theta_i -\pi/2
   = \widetilde{\theta}_i - \pi/2
   \label{EQ:symBSphases},
\eeq
at all wave numbers $k$. From this it follows that the delay difference is zero, $ \Delta \tau_\mathrm{BS} =0$, and the beamsplitters do not affect the phase averaging.  The effect of asymmetric beamsplitters is discussed in \secref{SEC:roleBS}.

Then, assuming symmetric beamsplitters and parameters typical of hot-electron experiments (see Appendix \ref{SEC:params}) we find the offset length to assume a value $l_\mathrm{offset} \approx 700$\,nm. In contrast, the effective single-particle coherence length is $\Sigma_l \approx 34$\,nm for a wavepacket of energetic width $\sigma_E=1$\,meV. Thus, it is essential that the construction of a hot-electron MZI  be such that the path difference satisfies $
    l_0 \approx   l_\mathrm{offset}
$ at the injection energy $E_0$ if interference is to be observed. It is perhaps worth noting that the offset length at high magnetic field is largely independent of the field strength because the dependency from the cyclotron frequency cancels with that of the velocity.

\section{Numerical results \label{sec:numericalResults}}

In this section, we evaluate the arrival time distribution of electrons by numerical integration of \eq{EQ:Pxt2} using the MZI phases of \secref{SEC:MZIphases} and parameters described in Appendix~\ref{SEC:params}.  
We take it that a typical experiment will look for oscillations as a function of injection energy with all other parameters fixed, and we thus recast the above results in terms of energy.
Since the predicted offset $l_\mathrm{offset}$ in \eq{EQ:loffset} depends on injection energy $E_0$ (through $v_0$), for a fixed path difference the maximum visibility condition $l_0 =  l_\mathrm{offset}$ defines an energy at which phase averaging effects vanish.  Denoting this energy as $E_\mathrm{peak}$, the phase averaging factor can be written as
\beq
  \mathcal{D}
  &=&
  \exp 
  \left\{
    -\frac{1}{2} 
    \rb{
      \frac{
       E_0 - E_\mathrm{peak}
      }{
      \Sigma_E
      }
    }^2    
  \right\}
  \label{eq:visibilityE}
\eeq
where
\beq
  \Sigma_E = \frac{ \hbar \omega_y^2E_\mathrm{peak}}{4 \omega_c \sigma_E}
  \label{EQ:VwidthE}
\eeq
is the energetic width of the visibility peak.  
For the results here we set $E_\mathrm{peak} = 100$\,meV such that we obtain an offset of $l_\mathrm{offset}\approx700$\,nm that we have taken into account in fixing the lengths of the MZI arms.

Initially we focus on the scenario where the beamsplitters are symmetric and from now on consider the two beamsplitters to be identical, such that e.g.
$
  T_1 = T_2 = T
$.
For concreteness we consider the Fertig-Halperin saddle model of a quantum point contact \cite{Fertig1987}, the transmission of which is
$
  T =  \left[1+\exp\left(-\pi \epsilon\right)\right]^{-1}
$,
with
$
 \epsilon = (E - V_0)/E^\mathrm{sad};
$
where 
$E^\mathrm{sad} = \hbar \omega_{\mathrm{BS}}^2/(2\omega_c)$ with $\hbar \omega_{\mathrm{BS}}$ an energy characterising the curvature of the saddle.
In Appendix \ref{SEC:BSphase}, we find the phase associated with transmission through the beamsplitter to be 
\beq
  \theta
  &=&
  - \frac{1}{2}\epsilon
  + \mathcal{X}_R \sqrt{\mathcal{X}_R^2+\epsilon}
  + \frac{1}{2} \epsilon \ln \rb{\frac{|\epsilon|}{2}}
  \nonumber\\
  &&
  - \epsilon g \rb{\mathcal{X}_R^2/\epsilon}
  + \mathrm{Im}
  \left[  
    \ln \Gamma\rb{\frac{1}{2} - \frac{i}{2}\epsilon}
  \right]
  \label{EQ:maintheta}
  ,
\eeq
where
$
  g(x) = 
  \ln
  \left|
    \sqrt{|x| + \mathrm{sgn}(x)} - \sqrt{|x|}
  \right|
$,
$
  \mathcal{X}_R = 
  \left[
    \rb{d/l_c}
    -
    \rb{2\epsilon^{(0)}l_c /d}
  \right]/\sqrt{8}
$,
and
$
 \epsilon^{(0)} = (E_0 - V_0)/E^\mathrm{sad}
$.
In this model, the size $d$ of the beamsplitter region is rather arbitrary, and here we set  $d = 3\sqrt{2} l_c$.  As explained in the appendix, the finiteness of this beamsplitter region means that the theory is only valid for energies
$
  |E - V_0| \le  
  \hbar \omega_{BS}^2 d^2
   / \rb{4\omega_c l_c^2} 
$; otherwise, a semi-classical trajectory of the electron of energy $E$ does not enter the beamsplitter region.
The other three beamsplitter phases are obtained through the relation (\ref{EQ:symBSphases}) in the symmetric case.

\begin{figure}[tb]
	\centering
	\includegraphics[width=\columnwidth,clip=true]{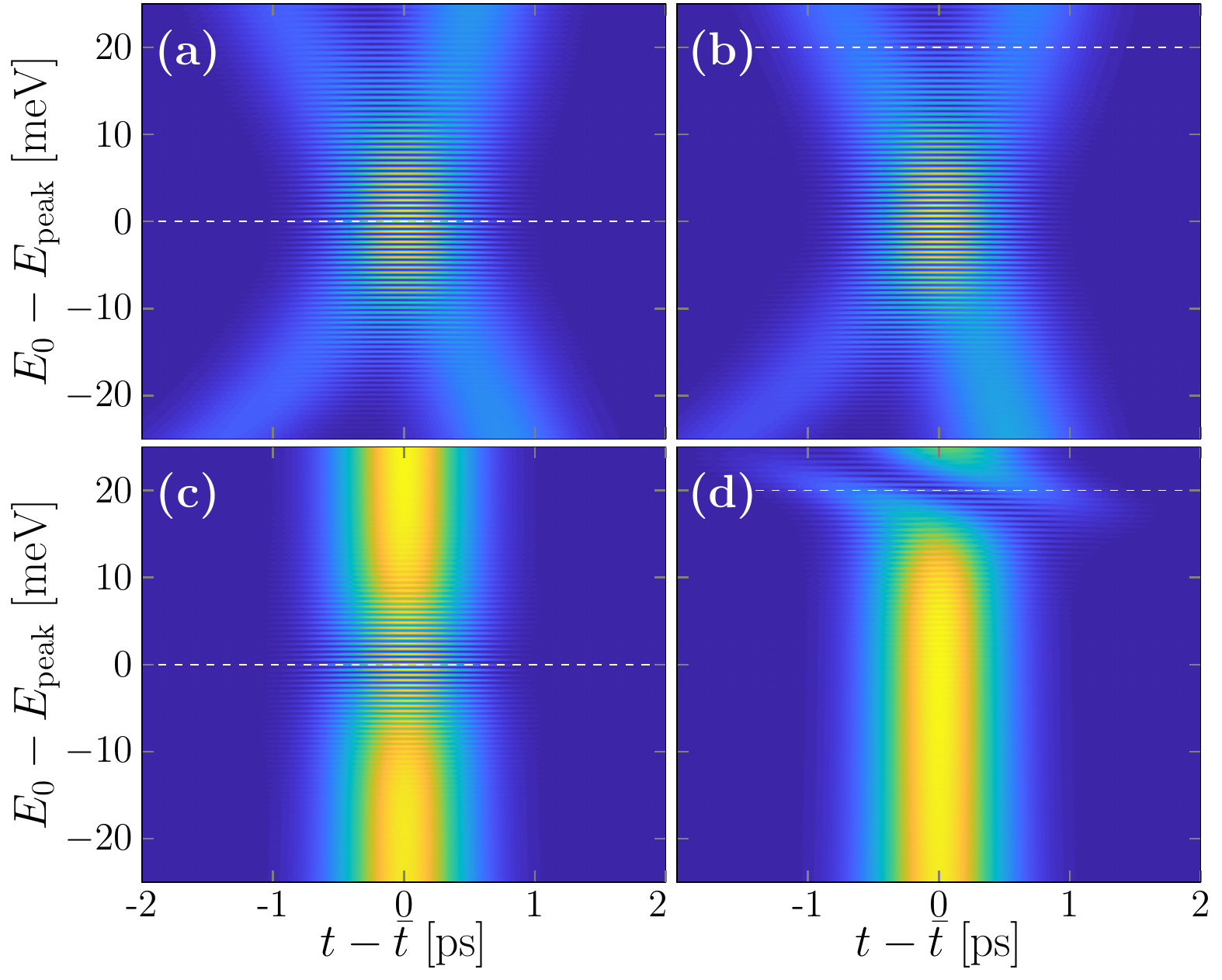}
	\caption{
    The arrival time distributions of electrons after they travel through an MZI as a function of time and injection energy $E_0$, calculated through the numerical evaluation of \eq{EQ:Pxt2}.  The arrival-time probability density is highest in the light/yellow regions, and lowest (tending to zero) in the dark/blue regions.
    The MZI path difference $l_0$ is set such that $ l_0 = -l_\mathrm{offset}$ at an energy of $E_\mathrm{peak} = 100$\,meV. Results are plotted about the mean arrival time $\overline{t}$ at each energy.
    Panels (a) and (b) show results for $\hbar \omega_\mathrm{BS} = 100$\,meV with $V_0=100$\,meV and $V_0=120$\,meV respectively (marked with horizontal lines).  With this beamsplitter width, interference is observed around $E_0 = E_\mathrm{peak}$ and the position of $V_0$ is unimportant.
    Panels (c) and (d) show the same but with $\hbar \omega_\mathrm{BS} = 20$\,meV.
    For this narrower beamsplitter, interference is observed around $V_0$ instead.
    Parameters as described in Appendix~\ref{SEC:params}.
	}
	\label{fig:sym_num_results}
\end{figure}

\begin{figure}[tb]
	\centering
	\includegraphics[width=\columnwidth]{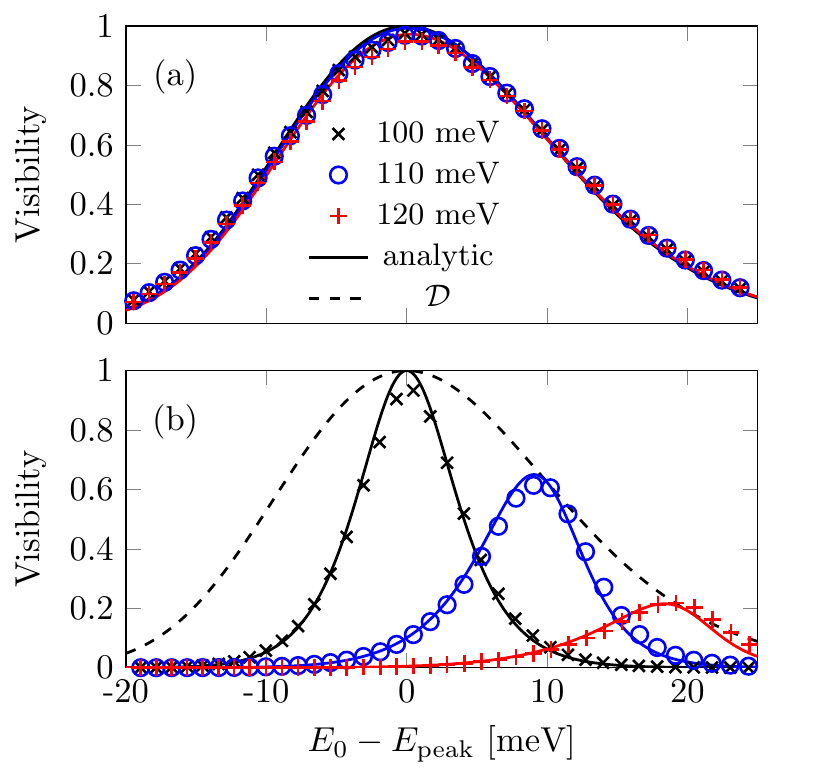}
	\caption{
    The MZI visibility as a function of injection energy $E_0$ for three values of the beamsplitter centre, $V_0 = 100$, $110$ and $120$\,meV.  Part (a) shows results for a beamsplitter width $\hbar \omega_\mathrm{BS} = 100$\,meV; part (b) for $\hbar \omega_\mathrm{BS} = 20$\,meV. Numerical results determined from \fig{fig:sym_num_results} are shown as symbols; the analytic expression of \eq{eq:full-expression} is shown as solid lines and the phase-averaging factor $\mathcal{D}$ of \eq{eq:visibility} is shown as a dashed line.
    }
	\label{fig:analytics_vs_numerics}
\end{figure}

Figure \ref{fig:sym_num_results} shows the numerical energy-dependent arrival time distribution,  plotted about the mean arrival time $\overline{t} = \int dt~ t~ \mathcal{A}(t)$.
Parts (a) and (b) of this figure show results for when the effective width of the beamsplitter transmission $E^\mathrm{sad}$ is large compared with the visibility width of \eq{EQ:VwidthE} ($\hbar \omega_\mathrm{BS} = 100$\,meV giving $E^\mathrm{sad} \approx 263$\,meV  compared with a width of $\Sigma_E \approx 10$\,meV).
%
%
In this case, both interferometer paths contribute to the arrival time distribution across the range shown and the picture captured by analytic expression \eq{EQ:PUxt} is very much born out here.  We see two lobes in the arrival time distribution that coincide when $E_0 = E_\mathrm{peak}$ such that interference takes place. The oscillation period is given by
$\delta E_0 \approx {\pi \hbar \omega_y^2}/\rb{2 \omega_c}\approx 0.6$\,meV 
here.
\fig{fig:analytics_vs_numerics} shows the visibility extracted from this numerical data (symbols), in comparison with the analytic result of \eq{eq:full-expression} (solid lines). For the wide beamsplitter case shown in \fig{fig:analytics_vs_numerics}a, we see that the visibility is hardly affected by moderate changes in the centre of the beamsplitter transmission $V_0$.  Thus we see the whole of the Gaussian visibility feature predicted by \eq{eq:visibilityE}.

In contrast, Figs.~\ref{fig:sym_num_results}c and d show results for a narrower beamsplitter with $\hbar \omega_\mathrm{BS} = 20$\,meV and $E^\mathrm{sad} = 10.52$\,meV, which is comparable with the visibility width $\Sigma_E$.  In this case, the appearance of fringes is heavily influenced by the beamsplitter transmission. For the upper and lower energy ranges of  \fig{fig:sym_num_results}c, only a single path of the MZI is traversed, and thus interference is restricted to a narrow range around $V_0$ where the product $TR$ is significantly different from zero.  The energy at which this occurs changes as we vary the beamsplitter centre $V_0$, as can be appreciated in \fig{fig:sym_num_results}d where $V_0 = 120$\,meV and the fringes occur at the top of the displayed energy range.
These changes are apparent in the visibility plots of \fig{fig:analytics_vs_numerics}b, where we see numerical visibility for three different values of $V_0$.
We note that the phase-averaging factor $\mathcal{D}$ (dashed line) acts as an approximate envelope for the maximum visibilities as $V_0$ is changed.

\section{Role of the beamsplitter phases \label{SEC:roleBS}}

The effect of the beamsplitter phases on the MZI properties can be captured by the two quantities: the mean delay
\beq
  \overline{\tau}_\mathrm{BS} = \frac{1}{2}v_0^{-1}\frac{d}{dk}\rb{\rho_1+\widetilde{\rho_2} + \theta_1+\theta_2},
\eeq
and the delay difference of \eq{eq:asymm_delays}.
In the symmetric case, the wavepacket delay given by the phase of
\eq{EQ:maintheta} is 
\beq
  \tau_\theta = \frac{\hbar}{2 E^\mathrm{sad}}
  \left\{
    \ln \frac{|\epsilon|}{2}
    -2 g \rb{\frac{\mathcal{X}^2}{\epsilon}}
    - \mathrm{Re}\, \psi\rb{\frac{1}{2} + \frac{i}{2} |\epsilon|}
  \right\}
  \nonumber\\
   \label{eq:BS_delay_sym}
\eeq
where $\psi$ is the digamma function. In this case, since all beamsplitter delay times are identical, we have $ \overline{\tau}_\mathrm{BS} =2 \tau_\theta $ and $\Delta \tau_\mathrm{BS}=0$. As noted above then, the beamsplitter plays no role in determining the interference, only in shifting the overall position of the arrival time distribution.

We now introduce an asymmetry into the action of the beamsplitters by considering them to be described by saddle potentials with a different curvature on either side. Details of this model are given in Appendix \ref{SEC:BSphase}.  Here it suffices to say that we assume both saddle to be alike and described by a parameter $\eta$ that gives the degree of asymmetry of the saddle, with $\eta = 0$ equivalent to the symmetric case above, and $\eta \to 1$ an extreme limit of asymmetry in which the beamsplitter all but closes on one side.
The energy range for which this model of the beamsplitter is valid is
\beq
  \frac{|E - V_0|}{\hbar} \le  
  \rb{\frac{d}{l_c}}^2
  \frac{\omega_{BS}^2}{2\omega_c} \cos^2 \left[\frac{\pi}{4}(\eta +1) \right]
  \label{EQ:ASYMenergyrange}.
\eeq
Thus, as the degree of asymmetry increases, the region of validity decreases, and goes to zero for $\eta\to1$.

With these asymmetric beamsplitters, there are two configurations: either the two directions of asymmetry are aligned, or they are opposite.
In the former case we have $\theta_1=\theta_2=\theta_\mathrm{sad}$, $\rho_1 = \rho_\mathrm{sad}$, and $\widetilde{\rho}_2 = \widetilde{\rho}_\mathrm{sad}$, where $\theta_\mathrm{sad}$ etc are the scattering phases of saddle potential as calculated in Appendix~\ref{SEC:BSphase}.  In this case one can show that the delay difference vanishes, $\Delta \tau_\mathrm{BS}=0$ as in the symmetric case. Thus, although the arrival time distribution will be modified by change in the transmission and reflection probabilities of the asymmetric model, the difference from the symmetric case is relatively minor. We note that this cancellation of $\Delta \tau_\mathrm{BS}$ does require the beamsplitters to be identical; small differences in parameters will lead to a small residual value for this quantity.

\begin{figure}[tb]
	\centering
	\includegraphics[width=\columnwidth,clip=true]{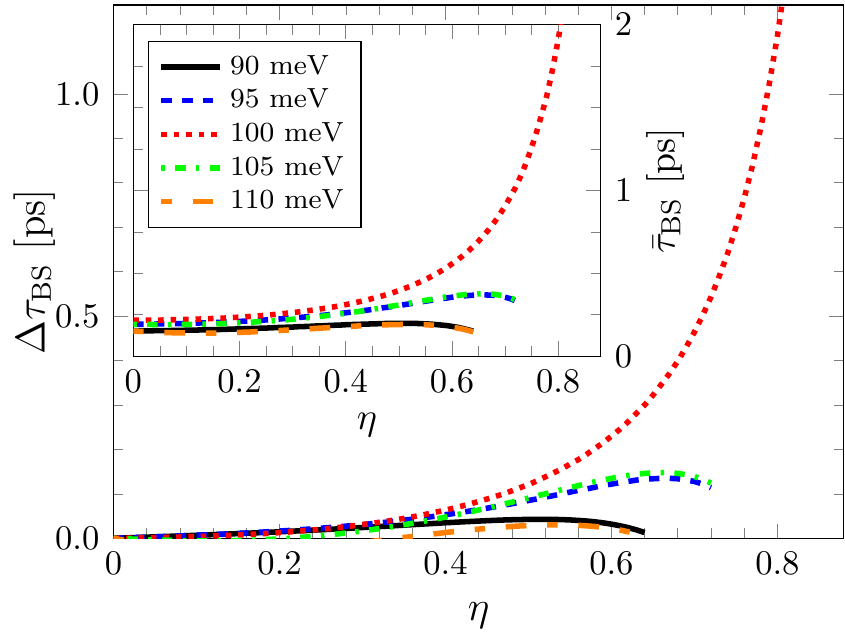}
	\caption{
    The main panel shows the beamsplitter contribution to the difference in delay times $\Delta \tau_\mathrm{BS}$ as a function of beamsplitter asymmetry parameter $\eta$ for various injection energies $E_0$.  The  inset shows the corresponding mean beamsplitter delay time  $\bar{\tau}_\mathrm{BS}$. The cut-off shown by these results arises from \eq{EQ:ASYMenergyrange}.
    The parameters are as in Appendix \ref{SEC:params}, with $V_0 = 100$\,meV and $\hbar \omega_\mathrm{BS} = 20$\,meV.
  }
	\label{fig:tau_bar_Delta_tau}
\end{figure}

Of greater interest here is the case when the beamsplitter asymmetries are oppositely aligned. Here we have $\theta_1=\theta_\mathrm{sad}$, $\theta_2=\widetilde{\theta}_\mathrm{sad}$, and $\rho_1 = \widetilde{\rho}_2 = \rho_\mathrm{sad}$.  In this case, the overall contribution of the beamsplitters to the travel-time difference does not vanish.
Indeed, 
\fig{fig:tau_bar_Delta_tau} shows the delay-time difference as a function of asymmetry for typical parameters and for several different values of $E_0$.  The mean delay time of the beamsplitters is also plotted.
We see that both these quantities have a very similar dependence on both $\eta$ and $E_0$, with the main difference being that $\Delta \tau_{BS}\to 0$ for $\eta \to 0$ whereas $\overline{\tau}_\mathrm{BS}$ tends to the finite value of \eq{eq:BS_delay_sym}.
For a given value of $\eta$, the delay difference is maximised when $E_0 = V_0$.  As $\eta$ increases, so does $\Delta \tau_\mathrm{BS}$ and, for $E_0 = V_0$ this time even diverges in the limit $\eta \to 1$ as the confinement on one side of the beamsplitters becomes flat.

The size of the delay difference determines the role that the beamsplitters play in the interference pattern.  Taking an example of $\eta = 1/2$ with $E_0 = V_0 = E_\mathrm{peak} =100$\,meV we obtain $\Delta \tau_\mathrm{BS} \approx 0.1$\,ps. Since this is small compared with temporal width of the wavepacket, $\sim 1ps$, this level of asymmetry will not significantly affect the observed interference. In contrast, For larger asymmetries, e.g. $\eta = 3/4$, we find  $\Delta \tau_\mathrm{BS} = 0.7$\,ps and since this is comparable with the temporal width, we can expect beamsplitter phases to be important here.

\begin{figure}[tb]
	\centering
	\includegraphics[width = 1 \columnwidth,clip=true]{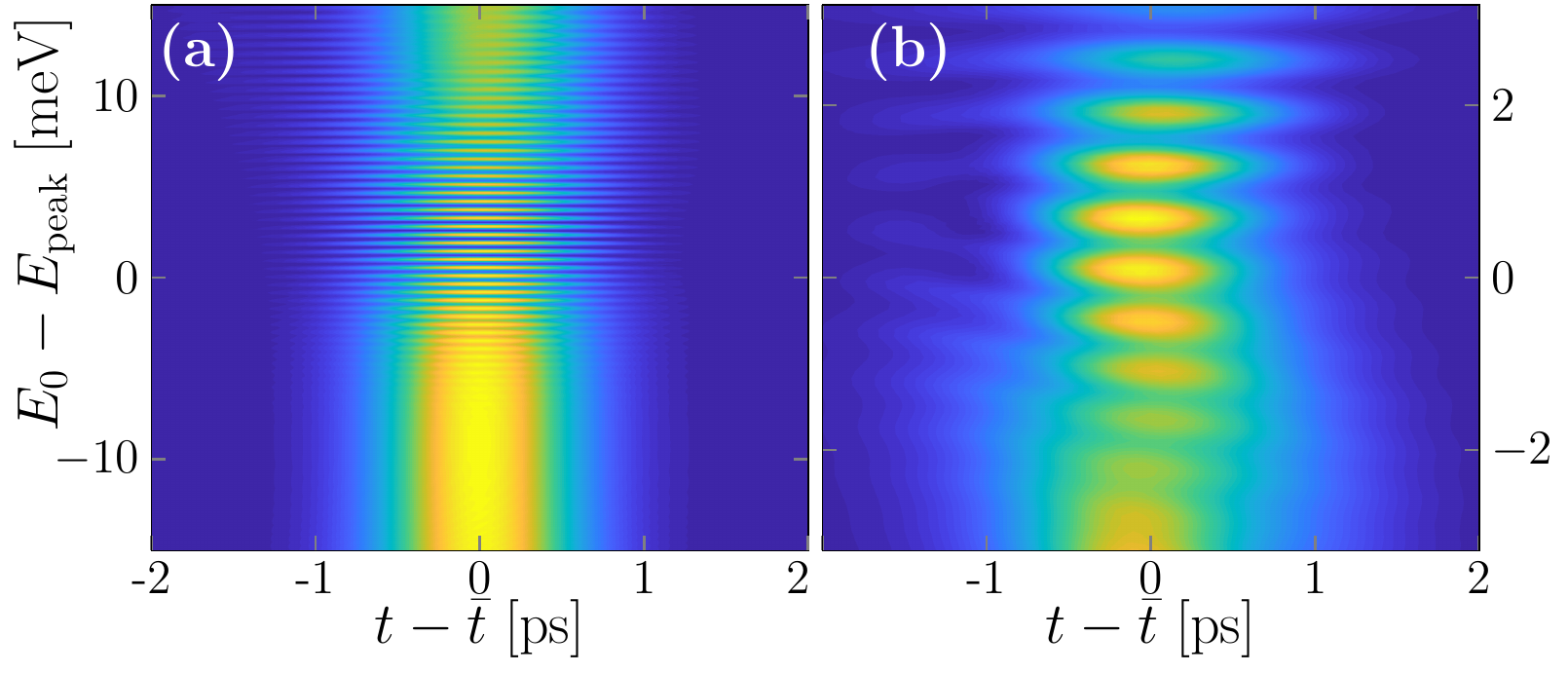}
	\caption{
	    As \fig{fig:sym_num_results}c but here with asymmetric beamsplitters with asymmetry parameter: (a) $\eta = 1/2$, and (b)  $\eta = 3/4$.
	  Increasing the asymmetry first shifts the position of maximum visibility [as in (a)] and then at higher levels [as in (b)] modifies the interference pattern. The energy ranges used here are compatible with keeping the majority of the wavepacket within the window defined by \eq{EQ:ASYMenergyrange}.
	  The arrival time distribution is greater in the lighter regions.
	}
	\label{fig:asymmetric_taushift}
\end{figure}

\fig{fig:asymmetric_taushift} shows the energy-dependent arrival time distribution of an electron for two values of asymmetry: (a)  $\eta = 1/2$ and (b) $\eta = 3/4$, with other parameters that match \fig{fig:sym_num_results}c (which may be thought of as the $\eta=0$ case in this sequence). In particular, the path difference $l_0$ is the same as in the $\eta=0$ case.
\fig{fig:asymmetric_taushift}a shows that this level of asymmetry changes the arrival distribution very little, with the main effect being that the distribution becomes non-symmetric about $E_0 - V_0 = 0$. This stems from a corresponding asymmetry about $V_0$ in the transmission probability of the beamsplitters. 
Increasing $\eta$ further, as in \fig{fig:asymmetric_taushift}b, we observe that the interference pattern becomes significantly modified indicating that the beamsplitter phases here are playing a significant role in determining the details of the interference.  We note that, for wider saddles [e.g. $\hbar \omega_\mathrm{BS} = 100$\,meV such as in \fig{fig:sym_num_results}a], the magnitude of $\Delta \tau^\mathrm{BS}$ is substantially less than observed here and the effect on the arrival time distributions for the same level of asymmetry is negligible.

\section{Discussion \label{sec:discussion}}

We have presented a wavepacket picture of the interference in a hot electron MZI, and focused on the energy-dependent arrival time distribution, as can be read out in experiment \cite{Fletcher2013,Waldie2015}.
The arrival time distributions clearly show two lobes, corresponding to the two paths travelled by the electrons, and the appearance of interference between the lobes depends on them arriving with overlap at the second beamsplitter. Therefore the travel time of the partial wavepackets relative to one another is of critical importance.  

We have shown here that for electrons an extension to the usual photon-optics approach of calculating the phase delays along the arms is required because, in the quantum Hall effect the position of the edge channels is dependent on energy, and this affects both dynamical and Aharonov-Bohm phases.
By taking this effect into account we have found that the peak interference occurs not at a path difference of zero, as it would in optics and has been discussed in previous analytic treatment of electronic phase averaging \cite{Chung2005,Haack2011,Beggi2015}, but rather at a finite offset.
We note that, although in their analytic calculation, Beggi \etal \cite{Beggi2015} derived a visibility with a zero offset, in their full simulations of electron transit through an MZI, they did find evidence of the maximum visibility being located at path difference away from zero, and similar in size to that predicted by \eq{EQ:loffset} for parameters appropriate to that calculation.

A key feature of the offset is that its size is proportional to the electron velocity. 
For electrons injected near the Fermi surface (e.g $E_0=5$\,meV), we find $l_\mathrm{offset} \sim 100$\,nm.  This value is small compared with typical interferometer dimensions. It is also small relative to single-particle coherence lengths for cold electrons, particularly in the case of mesoscopic-capacitor experiments where the coherence length is of the order of a micron \cite{Haack2011}. This  perhaps explains why this effect has not been considered previously.
In contrast, for hot electrons the value of $l_\mathrm{off}$ becomes closer to 1\,$\mu$m.  Clearly such a significant offset will make a difference to MZI design, with the geometry of experiments needing to be tailored to a specific injection energy to ensure optimal visibility.

We have also considered the role of beamsplitters in determining interference patterns. For energetically wide beamsplitters, the size of visibility features is essentially determined by the effective single-particle coherence length $\Sigma_l$.  Away from this limit, the transmission properties of the beamsplitters become important for determining the range of energy over which oscillations can be observed.
Concerning the contribution of the beamsplitters to the travel times and hence to phase averaging, if the beamsplitters are symmetric or if they are asymmetric but alike and aligned, then they only contribute an overall delay to the arrival times, and no loss of visibility is observed.  On the other hand, opposing asymmetries in the beamsplitters can affect both the quality and the structure of the interference patterns.  However, we found that this only becomes relevant at rather high degrees of asymmetry.

Finally we note that, despite the changes outlined in this work, the important conclusion from photon optics remains that, by setting the path length correctly, phase averaging can be effectively ``switched off''.  In this case other decoherence mechanisms, such as phonon emission and electron-electron interaction will then dominate \cite{Clark2020}.

\acknowledgments
This research was supported by EPSRC Grant No. EP/P034012/1. 
SR was supported by the María de Maeztu Program for units of Excellence in R\&D (MDM-2017-0711).
LAC also acknowledges support from the Foundation for Polish Science within the ``Quantum Optical Technologies'' project carried out within the International Research Agendas programme co-financed by the European Union under the European Regional Development Fund.
HSS was supported by the Korea NRF (SRC Center for Quantum Coherence in Condensed Matter, Grant No. 2016R1A5A1008184).
MK was supported by the UK Depart-
ment for Business, Energy, and Industrial Strategy, and
by the European Metrology Programme
for Innovation and Research (EMPIR).  This project 17FUN04 SEQUOIA has received funding from the EMPIR programme co-financed by the Participating States and from the European Union’s Horizon 2020 research and innovation programme.

\appendix

\section{Numerical parameters\label{SEC:params}}

The results discussed in this work were obtained with magnetic field strength of $B=11$\,T, transverse confinement of $\hbar \omega_y = 2.7$\,meV \cite{Johnson2018} and starting wavepacket width of $\sigma_E = 1$\,meV.
To determine the offset, we chose a peak energy of $E_\mathrm{peak} = 100$\,meV, which corresponds to an assumed injection energy in hot-electron experiments. With this fixed, we chose an MZI geometry with lengths (as defined in \fig{fig:mzi}) of
$X = 5\,\mu$\,m, $Y_L =  2\,\mu$\,m, and $Y_U = Y_L + \frac{1}{2} l_\mathrm{offset}(E_\mathrm{peak}) $, where we calculate the offset in the symmetric-beamsplitter limit (i.e. $\Delta \tau_\text{BS} = 0$ in \eq{EQ:loffset}) and obtain $l_\mathrm{offset}\approx 706.5$\,nm.  Finally we set the detector position as $x_\mathrm{D} = 0$ as it plays no important role here, and used the effective mass  $m_e^* = 0.067 m_e$ for GaAs.

\section{Scattering properties of an asymmetric saddle \label{SEC:BSphase}}

\begin{figure}[tb]
  \centering
  \includegraphics[width=0.6\columnwidth,clip=true]{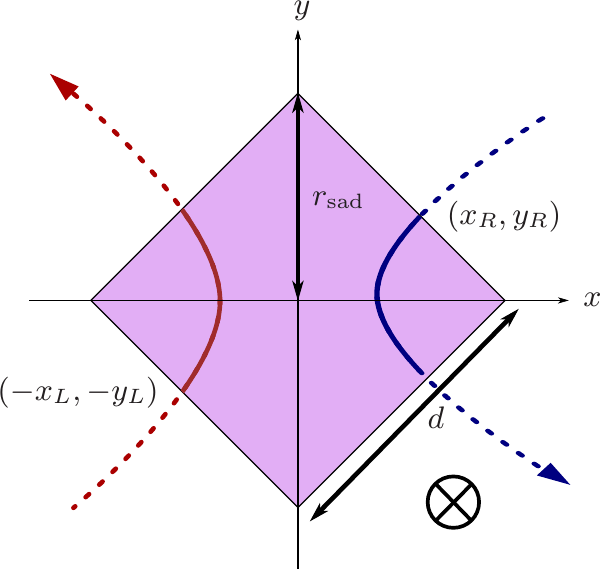}
  \caption{
    Beamsplitter region of saddle-point constriction. The asymmetric saddle potential of \eq{eq:U-saddle} is centred around the point $(x,y) = (0,0)$, and the region defined as the saddle region is indicated by the purple square. Once rotated this matches up with the beamsplitters of \fig{fig:mzi} with $d=\sqrt{2}r_\mathrm{sad}$. Indicative trajectories are shown on the left and the right, marked with the coordinates of where they enter the beamsplitter region.
  }
  \label{fig:saddle}
\end{figure}

Here we extend the work of Ref.~\cite{Fertig1987} to calculate the transmission properties of the  asymmetric saddle-point constriction with potential
\begin{equation}
  \label{eq:U-saddle}
  V_\text{sad}(x,y) =
  \begin{cases}
    V_0 +\frac{1}{2} m_e^* \omega_{\text{L}}^2 \rb{y^2- x^2} & \text{for } x<0
    ;
    \\
    V_0   +\frac{1}{2} m_e^* \omega_{\text{R}}^2 \rb{y^2- x^2} & \text{for } x > 0
    ,   
  \end{cases}
\end{equation}
where $\hbar \omega_{L/R}$ are the confinement energies either side of the saddle.
We first define the beamsplitter region as the region enclosed by a square of half-diagonal length $r_{\text{sad}}$ centred at the saddle point, see Fig.~\ref{fig:saddle}.
In the symmetric case we set $\omega_\mathrm{L}=\omega_\mathrm{R}=\omega_{BS}$, and the L/R subscript can be dropped.  For the asymmetric parametrisation discussed in the main paper, we set 
\beq
  \omega_\mathrm{L} &=& \sqrt{2}\omega_\mathrm{BS} \cos\left[\frac{1}{4}\pi(1+\eta)\right]
  \nonumber\\
  \omega_\mathrm{R} &=& \sqrt{2}\omega_\mathrm{BS} \sin\left[\frac{1}{4}\pi(1+\eta)\right]
\eeq
Here $0 \le \eta \le 1$ is a parameter describing the left-right asymmetry of the beamsplitter chosen such that mean width  
$
\frac{1}{2} \rb{E_\mathrm{L}^\mathrm{sad}+E_\mathrm{R}^\mathrm{sad}}
= \hbar \rb{\omega_{L}^2 + \omega_{R}^2}/\rb{4\omega_c}
= \hbar\omega_{BS}^2 / \rb{2\omega_c}
$ is independent of $\eta$ and compares directly with the expression $E^{\mathrm{sad}}$ in the symmetric case.

We consider an electron of energy $E$ (measured from the subband bottom) in the lowest Landau level~\cite{ryu:ultrafast}, whose guide centre enters in the beamsplitter region at position of $x=-x_{\text{L}}(E) ,y= -y_{\text{L}}(E)$ (see Fig.~\ref{fig:saddle}) and exits out of the region at $x= x_{\text{R}}(E), y=-y_{\text{R}}(E)$ for a transmission event and $x= -x_{\text{L}}(E), y=y_{\text{L}}(E)$ for a reflection event.
The approach followed here is only well defined  when the classical trajectory of the incident electron enters and exits the beamsplitter region.
This is satisfied when $\text{max} [V_{\text{sad}}(-r_{\text{sad}}, 0), V_{\text{sad}}(r_{\text{sad}}, 0)] < E < \text{min}[V_{\text{sad}}(0^{-}, -r_{\text{sad}}), V_{\text{sad}}(0^{+}, -r_{\text{sad}})] $, equivalently
\begin{equation}
  \label{eq:E-range}
  - \frac{\text{min}(\omega^2_{\text{L}}, \omega^2_{\text{R}})}{2\omega_{c}}
  < \rb{\frac{E - V_0}{\hbar}}\big(\frac{l_c}{r_{\text{sad}}}\big)^2 
  <  \frac{\text{min}(\omega^2_{\text{L}}, \omega^2_{\text{R}}) }{2 \omega_{c}}
  .
\end{equation}
To calculate the scattering properties of this model, we apply the transformation described in Ref.\,\cite{Fertig1987} for each region $x<0$ and $x>0$.
The transformed system of the one-dimensional Hamiltonian
\begin{equation}
  \label{eq:H1}
  H_1 =
  \begin{cases}
    V_0 +E_\mathrm{L}^\mathrm{sad} (\mathcal{P}^2 -\mathcal{X}^2) & \text{for } \mathcal{X}<0  \\
    V_0 +E_\mathrm{R}^\mathrm{sad} (\mathcal{P}^2 -\mathcal{X}^2) & \text{for } \mathcal{X}>0  
  \end{cases}
\end{equation}
describes the motion of the guide-centre energy in a strong magnetic field.
Here $E_\mathrm{L}^\mathrm{sad} =\hbar \omega_{\text{L}}^2 /(2\omega_c)$ and
$E_\mathrm{R}^\mathrm{sad} =\hbar \omega_{\text{R}}^2 /(2\omega_c)$. 
The dimensionless coordinate $\mathcal{X}= x / l_c$ describes the position of the guide center and $\mathcal{P}$ in the Hamiltonian $H_1$ is the canonical conjugate of $\mathcal{X}$, namely $[\mathcal{X}, \mathcal{P}]=i$.

The boundary conditions at $\mathcal{X}=0$ for an energy eigenstate $\psi_E$ of Hamiltonian $H_1$ are~\cite{levy1995position},
\begin{equation}
  \label{eq:bc}
  \begin{aligned}
    \psi_E(\mathcal{X}=0^{-}) &= \psi_E(\mathcal{X}=0^{+})  \\
    E_\mathrm{L}^\mathrm{sad} \frac{\partial \psi_E}{\partial \mathcal{X}}\Big|_{\mathcal{X}=0^{-}}
    &=E_\mathrm{R}^\mathrm{sad} \frac{\partial \psi_E}{\partial \mathcal{X}}\Big|_{\mathcal{X}=0^{+}}
  \end{aligned}
\end{equation}
Using the even and odd solutions for each $\mathcal{X}<0$ and $\mathcal{X}>0$ part of $H_1$~\cite{Fertig1987}, we have
\begin{equation}
  \label{eq:psi-asym}
  \begin{aligned}
    \psi_E(\mathcal{X}<0) &= A \phi_{\text{e}} (\mathcal{X}; \epsilon_\text{L}) +B \phi_{\text{o}}(\mathcal{X}; \epsilon_\text{L});    
    \\
    \psi_E(\mathcal{X}>0) &= C \phi_{\text{e}} (\mathcal{X}; \epsilon_\text{R}) +D \phi_{\text{o}}(\mathcal{X}; \epsilon_\text{R})
    ,
  \end{aligned}
\end{equation}
where $\epsilon_\sigma \equiv (E-V_0)/E_\sigma^\mathrm{sad}$; $\sigma\in\left\{\text{L},\text{R}\right\}$, and
\begin{equation}
  \label{eq:sol_eo}
  \begin{aligned}
    \phi_{\text{e}} (\mathcal{X};\epsilon ) &= e^{-i \mathcal{X}^2/2}
    F\big(\frac{1}{4} + \frac{1}{4}i \epsilon \big|\frac{1}{2} \big| i \mathcal{X}^2 \big), \\
    \phi_{\text{o}} (\mathcal{X}; \epsilon) &= \mathcal{X} e^{-i \mathcal{X}^2/2}
    F\big(\frac{3}{4} + \frac{1}{4} i \epsilon \big| \frac{3}{2} \big| i \mathcal{X}^2 \big),
  \end{aligned}
\end{equation}
with $F(a|b|c)$ a confluent hypergeometric function of 1st kind~\cite{Abramowitz1972}.
Substituting Eqs.~(\ref{eq:psi-asym}) and ~(\ref{eq:sol_eo}) into Eq.~(\ref{eq:bc}), we obtain
\begin{equation}
  \label{eq:bc-coeff}
  \begin{aligned}
    A F\big(\frac{1}{4} +\frac{1}{4} i \epsilon_{\text{L}} \big| \frac{1}{2} \big| 0)
    &= C F\big(\frac{1}{4} +\frac{1}{4} i \epsilon_{\text{R}} \big| \frac{1}{2} \big| 0); \\
    B F\big(\frac{3}{4} +\frac{1}{4}i \epsilon_{\text{L}} \big| \frac{3}{4} \big| 0)
    &= \frac{E_\mathrm{R}^\mathrm{sad}}{E_\mathrm{L}^\mathrm{sad}} D F\big(\frac{3}{4} +\frac{1}{4}i \epsilon_{\text{R}} \big| \frac{3}{4} \big| 0).
  \end{aligned}
\end{equation}
Using $F(a|b|0) =1$~\cite{Abramowitz1972}, we find $A = C$ and $B = \frac{E_\mathrm{R}^\mathrm{sad}}{E_\mathrm{L}^\mathrm{sad}} D $.

Now we impose a condition that $\psi_{E}$ is a scattering state generated by an incoming wave from $\mathcal{X}<0$.
The asymptotic approximations for $\phi_\text{e}$ and $\phi_\text{o}$ at $|\mathcal{X}| \gg 1$\cite{Fertig1987} are,
\begin{align}
\label{sol-eo-asymp}
  \phi_{\text{e}} (\mathcal{X}; \epsilon) &\rightarrow 
  \frac{\Gamma(\frac{1}{2})}{\Gamma(\frac{1}{4} + \frac{1}{4} i \epsilon)}
  e^{- i \frac{\pi}{8} - \frac{\pi}{8} \epsilon }
  |\mathcal{X}|^{-\frac{1}{2} + i \frac{1}{2}\epsilon} e^{i \mathcal{X}^2/2} 
  \nonumber\\
  &
  +\frac{\Gamma(\frac{1}{2})}{\Gamma(\frac{1}{4} - \frac{1}{4} i \epsilon)}
  e^{ i \frac{\pi}{8} - \frac{\pi}{8} \epsilon }
  |\mathcal{X}|^{-\frac{1}{2} - i \frac{1}{2}\epsilon} e^{-i \mathcal{X}^2/2}     \\
  \phi_{\text{o}} (\mathcal{X}; \epsilon) &\rightarrow  \mathcal{X}\Big[
  \frac{\Gamma(\frac{3}{2})}{\Gamma(\frac{3}{4} + \frac{1}{4} i \epsilon)}
  e^{- i \frac{3\pi}{8} - \frac{\pi}{8} \epsilon }
  |\mathcal{X}|^{-\frac{3}{2} + i \frac{1}{2}\epsilon} e^{i \mathcal{X}^2/2} 
  \nonumber\\
  &
  +\frac{\Gamma(\frac{3}{2})}{\Gamma(\frac{3}{4} - \frac{1}{4} i \epsilon)}
  e^{i \frac{3\pi}{8} - \frac{\pi}{8} \epsilon }
  |\mathcal{X}|^{-\frac{3}{2} - i \frac{1}{2}\epsilon} e^{-i \mathcal{X}^2/2} \Big]
\end{align}
The terms proportional to $e^{i\mathcal{X}^2/2}$ [$e^{-i\mathcal{X}^2/2}$] correspond to outgoing [incoming] waves.
Requiring that the incoming wave for $\mathcal{X}>0$ should cancel, we obtain the condition
\begin{equation}
  \label{inc-cond-asym}
  C \frac{\Gamma(\frac{1}{2})}{\Gamma(\frac{1}{4} -\frac{1}{4} i \epsilon_\text{R} ) } e^{i \pi/8}
  +D \frac{\Gamma(\frac{3}{2})}{\Gamma(\frac{3}{4} -\frac{1}{4} i \epsilon_\text{R}) } e^{i 3 \pi/8}  =0 .
\end{equation}
Applying an asymptotic approximation of Eq.~(\ref{sol-eo-asymp}) to Eq.~(\ref{eq:psi-asym}),
we obtain expressions for $\psi^{\text{(in)}}_{E}$ and $\psi^{\text{(out)}}_{E}$, incoming and outgoing parts of the scattering state, such that (for a coordinate $\mathcal{X}_0 \gg1$)
\beq  
    \frac{\psi^{\text{(out)}}_{E} (\mathcal{X}_0) }{\psi^{\text{(in)}}_{E} (-\mathcal{X}_0)}
    &=& e^{-\frac{\pi}{8} (\epsilon_\text{R}-\epsilon_\text{L}) } \mathcal{X}_0^{\frac{i}{2} (\epsilon_\text{L} +\epsilon_\text{R}) }  e^{ i \mathcal{X}_0^2}
    \label{eq:t-asymp}\\
    && \times
    \frac{ e^{-i \frac{\pi}{4}} \frac{\Gamma(\frac{1}{4}- \frac{1}{4} i \epsilon_\text{R}) }{\Gamma(\frac{1}{4}+ \frac{1}{4} i \epsilon_\text{R}) }
    -e^{-i \frac{3\pi}{4}} \frac{\Gamma(\frac{3}{4}- \frac{1}{4} i \epsilon_\text{R}) }{\Gamma(\frac{3}{4}+ \frac{1}{4} i \epsilon_\text{R}) } }
    { \frac{\Gamma(\frac{1}{4}- \frac{1}{4} i \epsilon_\text{R}) }{\Gamma(\frac{1}{4}- \frac{1}{4} i \epsilon_\text{L}) }
    +\frac{E_\mathrm{R}^\mathrm{sad}}{E_\mathrm{L}^\mathrm{sad}}\frac{\Gamma(\frac{3}{4}- \frac{1}{4} i \epsilon_\text{R}) }{\Gamma(\frac{3}{4} - \frac{1}{4} i \epsilon_\text{L}) } }
    \nonumber\\
    \frac{\psi^{\text{(out)}}_{E}(-\mathcal{X}_0) }{ \psi^{\text{(in)}}_{E}(-\mathcal{X}_0) };
    &=&  
    \mathcal{X}_0^{i \epsilon_\text{L} }  e^{ i \mathcal{X}_0^2}  
    \label{eq:r-asymp}\\
    && \times
    \frac{ e^{-i \frac{\pi}{4}} \frac{\Gamma(\frac{1}{4}- \frac{1}{4} i \epsilon_\text{R}) }{\Gamma(\frac{1}{4}+ \frac{1}{4} i \epsilon_\text{L}) }
    + \frac{E_\mathrm{R}^\mathrm{sad}}{E_\mathrm{L}^\mathrm{sad}} e^{-i \frac{3\pi}{4}} \frac{\Gamma(\frac{3}{4}- \frac{1}{4} i \epsilon_\text{R}) }{\Gamma(\frac{3}{4}+ \frac{1}{4} i \epsilon_\text{L}) } }
    { \frac{\Gamma(\frac{1}{4}- \frac{1}{4} i \epsilon_\text{R}) }{\Gamma(\frac{1}{4}- \frac{1}{4} i \epsilon_\text{L}) }
    +\frac{E_\mathrm{R}^\mathrm{sad}}{E_\mathrm{L}^\mathrm{sad}}\frac{\Gamma(\frac{3}{4}- \frac{1}{4} i \epsilon_\text{R}) }{\Gamma(\frac{3}{4} - \frac{1}{4} i \epsilon_\text{L}) } }
    \nonumber.
\eeq

\subsection{Phases and delay times}

To make use of the expressions derived above, we express quantities of interest as the difference of two terms evaluated asymptotically. Thus, we write the transmission and reflection phases as 
\beq
    \theta = \lim_{\mathcal{X}_0 \rightarrow \infty}
    [\theta^{\text{tot}} - \theta^{\text{ext}}]
  ;\qquad
     \rho = \lim_{\mathcal{X}_0 \rightarrow \infty}
    [\rho^{\text{tot}} - \rho^{\text{ext}}].
    \nonumber
\eeq
In these expressions the first term is the total phase accumulated in travelling from points $\mathcal{X} = -\mathcal{X}_0 $ to $\mathcal{X} =  \pm \mathcal{X}_0$ (upper and lower signs describe transmission and reflection respectively) far from the beamsplitter,  and the second terms represent the phase accumulated along the parts of these paths that are external to the saddle region.
The simplification that this brings is that, outside of the beamsplitter region, an electron wavepacket moves along a trajectory that be be determined semi-classically with an error at most $\exp[-\pi r_{\text{sad}}^2/l_c^2]  $ \cite{Fertig1987}.  Thus $\theta^{\text{ext}}$ and $\rho^{\text{ext}}$ may be evaluated semi-classically.

The quantum mechanical contributions to the above are then evaluated as
\beq
  \theta^\mathrm{tot} &=&
   \text{Im} 
   \left[
    \ln 
    \rb{
      \frac{\psi^{\text{(out)}}_{E}(\mathcal{X}_0)}{\psi^{\text{(in)}}_{E}(-\mathcal{X}_0)} 
      }
    \right]
  ;\nonumber\\
  \rho^\mathrm{tot} &=&
  \text{Im}
   \left[
    \ln 
    \rb{
      \frac{\psi^{\text{(out)}}_{E}(-\mathcal{X}_0)}{\psi^{\text{(in)}}_{E}(-\mathcal{X}_0)}
    }
   \right]
  \nonumber
  ,
\eeq
using the asymptotic results of \eq{eq:t-asymp} and \eq{eq:r-asymp}.
Meanwhile, the external phases evaluated semi-classically read
\beq
    \theta^{\text{ext}}( E)
    &= \int_{-\mathcal{X}_0}^{-\mathcal{X}_{\text{L}}(E)} d\mathcal{X} \sqrt{ \frac{E-V_0}{E_\mathrm{L}^\mathrm{sad}}+\mathcal{X}^2}
    \nonumber\\
    &~~ 
    +\int_{\mathcal{X}_{\text{R}}(E)}^{\mathcal{X}_0} d\mathcal{X} \sqrt{ \frac{E-V_0}{E_\mathrm{R}^\mathrm{sad}}+\mathcal{X}^2}
    ;
    \\
    \rho^{\text{ext}}( E)
    &=2 \int_{-\mathcal{X}_0}^{-\mathcal{X}_{\text{L}}(E)} d\mathcal{X} \sqrt{ \frac{E-V_0}{E_\mathrm{L}^\mathrm{sad}}+\mathcal{X}^2}
    .
  \eeq
These are expressed in terms of the coordinates 
\begin{equation}
  \label{eq:X-ent-exit}
    \mathcal{X}_{\sigma} (E)
    = 
    \frac{x_\sigma}{l_c}
    = 
    \frac{1}{2} \frac{l_c}{r_{\text{sad}}}
    \left[ \left(\frac{r_{\text{sad}}}{l_c}\right)^2 -  \frac{E-V_0}{E^\mathrm{sad}_{\sigma}} \right]
\end{equation}
Evaluating the integrals and taking the $\mathcal{X}_0 \to \infty$ limit,  we find 
\begin{equation}
  \begin{aligned}
    \theta^{\text{ext}}
    &=\sum_{\sigma=\text{L},\text{R}}\Big[\frac{1}{4}\epsilon_\sigma -\frac{1}{2} \mathcal{X}_{\sigma}(E) \sqrt{\mathcal{X}^2_{\sigma}(E) +\epsilon_{\sigma}}
    +\frac{1}{2} \mathcal{X}_0^2
    \\
    &
    -\frac{1}{4} \epsilon_{\sigma} \ln |\epsilon_{\sigma}| + \frac{1}{2} \epsilon_{\sigma} \ln (2\mathcal{X}_0)
    +\frac{1}{2} \epsilon_{\sigma} g(\mathcal{X}_{\sigma}(E)^2/\epsilon_{\sigma})\Big]
    ;
    \\
    \rho^{\text{ext}}
    &=\frac{1}{2}\epsilon_{\text{L}}  - \mathcal{X}_{\text{L}}(E) \sqrt{\mathcal{X}^2_{\text{L}}(E) +\epsilon_{\text{L}}}
    + \mathcal{X}_0^2
    - \frac{1}{2}\epsilon_{\text{L}} \ln |\epsilon_{\text{L}}|
    \\
    &
    + \epsilon_{\text{L}} \ln (2\mathcal{X}_0)
    + \epsilon_{\text{L}} g(\mathcal{X}_{\text{L}}(E)^2/\epsilon_{\text{L}})
    ,
  \end{aligned}
\end{equation}
with $
  g(x) = 
  \ln
  \left|
    \sqrt{|x| + \mathrm{sgn}(x)} - \sqrt{|x|}
  \right|
$, such that the complete phases read
\begin{widetext}
\begin{align}
  \theta(E)
  &=\sum_{\sigma=\text{L},\text{R}}\Big[-\frac{1}{4}\epsilon_{\sigma}
  +\frac{1}{2} \mathcal{X}_{\sigma}(E) \sqrt{\mathcal{X}^2_{\sigma}(E) +\epsilon_{\sigma}}
  +\frac{1}{4} \epsilon_{\sigma} \ln \frac{|\epsilon_{\sigma}|}{4}
  -\frac{1}{2} \epsilon_{\sigma} g(\mathcal{X}_{\sigma}(E)^2/\epsilon_{\sigma})\Big]
  \nonumber\\ 
  & \qquad
  +\text{arg} \Big[
  e^{-i \frac{\pi}{4}} \frac{\Gamma(\frac{1}{4}- \frac{1}{4} i \epsilon_\text{R}) }{\Gamma(\frac{1}{4}+ \frac{1}{4} i \epsilon_\text{R}) }
  -e^{-i \frac{3\pi}{4}} \frac{\Gamma(\frac{3}{4}- \frac{1}{4} i \epsilon_\text{R}) }{\Gamma(\frac{3}{4}+ \frac{1}{4} i \epsilon_\text{R}) } \Big]
  - \text{arg}\Big[\frac{\Gamma(\frac{1}{4}- \frac{1}{4} i \epsilon_\text{R}) }{\Gamma(\frac{1}{4}- \frac{1}{4} i \epsilon_\text{L}) }
    +\frac{E_\mathrm{R}^\mathrm{sad}}{E_\mathrm{L}^\mathrm{sad}}\frac{\Gamma(\frac{3}{4}- \frac{1}{4} i \epsilon_\text{R}) }{\Gamma(\frac{3}{4} - \frac{1}{4} i \epsilon_\text{L}) }  \Big]
\label{eq:theta}\\
  \rho(E)
  &=-\frac{1}{2} \epsilon_{\text{L}} + \mathcal{X}_{\text{L}}(E) \sqrt{\mathcal{X}^2_{\text{L}}(E) +\epsilon_{\text{L}}}
  + \frac{1}{2}\epsilon_{\text{L}} \ln \frac{|\epsilon_{\text{L}}|}{4}
  - \epsilon_{\text{L}} g(\mathcal{X}_{\text{L}}(E)^2/\epsilon_{\text{L}})
  \nonumber\\ 
  &\qquad
      + \text{arg}\Big[ 
  e^{-i \frac{\pi}{4}} \frac{\Gamma(\frac{1}{4}- \frac{1}{4} i \epsilon_\text{R}) }{\Gamma(\frac{1}{4}+ \frac{1}{4} i \epsilon_\text{L}) }
  + \frac{E_\mathrm{R}^\mathrm{sad}}{E_\mathrm{L}^\mathrm{sad}}e^{-i \frac{3\pi}{4}} \frac{\Gamma(\frac{3}{4}- \frac{1}{4} i \epsilon_\text{R}) }{\Gamma(\frac{3}{4}+ \frac{1}{4} i \epsilon_\text{L}) }\Big]
   -\text{arg} \Big[\frac{\Gamma(\frac{1}{4}- \frac{1}{4} i \epsilon_\text{R}) }{\Gamma(\frac{1}{4}- \frac{1}{4} i \epsilon_\text{L}) }
    +\frac{E_\mathrm{R}^\mathrm{sad}}{E_\mathrm{L}^\mathrm{sad}}\frac{\Gamma(\frac{3}{4}- \frac{1}{4} i \epsilon_\text{R}) }{\Gamma(\frac{3}{4} - \frac{1}{4} i \epsilon_\text{L}) }  \Big]
  \label{eq:rho}
\end{align}
The corresponding delay times are obtained by differentiating [as in \eq{EQ:firstdelaysbydiff}] to give
\begin{align}
  \tau_\theta (E) 
  &=
  \frac{\hbar}{4 E_\mathrm{L}^\mathrm{sad}} \ln ( |\epsilon_{\text{L}}|/4)
  -\frac{\hbar}{2 E_\mathrm{L}^\mathrm{sad}} g\big( \frac{\mathcal{X}_{\text{L}}^2(E)}{\epsilon_{\text{L}}}\big)
  +\frac{\hbar}{4 E_\mathrm{R}^\mathrm{sad}} \ln ( |\epsilon_{\text{R}}|/4)
  -\frac{\hbar}{2 E_\mathrm{R}^\mathrm{sad}} g\big( \frac{\mathcal{X}_{\text{R}}^2(E)}{\epsilon_{\text{R}}}\big) 
  \nonumber\\
  & \qquad + \hbar\, \text{Im} \frac{\partial}{\partial E} \ln \Big[
  e^{-i \frac{\pi}{4}} \frac{\Gamma(\frac{1}{4}- \frac{1}{4} i \epsilon_\text{R}) }{\Gamma(\frac{1}{4}+ \frac{1}{4} i \epsilon_\text{R}) }
  -e^{-i \frac{3\pi}{4}} \frac{\Gamma(\frac{3}{4}- \frac{1}{4} i \epsilon_\text{R}) }{\Gamma(\frac{3}{4}+ \frac{1}{4} i \epsilon_\text{R}) } \Big] 
  - \hbar \,\text{Im} \frac{\partial}{\partial E} \ln \Big[
  \frac{\Gamma(\frac{1}{4}- \frac{1}{4} i \epsilon_\text{R}) }{\Gamma(\frac{1}{4}- \frac{1}{4} i \epsilon_\text{L}) }
  +\frac{E_\mathrm{R}^\mathrm{sad}}{E_\mathrm{L}^\mathrm{sad}}\frac{\Gamma(\frac{3}{4}- \frac{1}{4} i \epsilon_\text{R}) }{\Gamma(\frac{3}{4} - \frac{1}{4} i \epsilon_\text{L}) }
  \Big]
  ;
  \label{eq:tau-dT-asym}
  \\
  \tau_\rho(E) &= 
  \frac{\hbar}{2 E_\mathrm{L}^\mathrm{sad}} \ln ( |\epsilon_{\text{L}}|/4)
  -\frac{\hbar}{ E_\mathrm{L}^\mathrm{sad}} g\big( \frac{\mathcal{X}_{\text{L}}^2(E)}{\epsilon_{\text{L}}}\big)
  + \hbar\, \text{Im} \frac{\partial}{\partial E} \ln \Big[
  e^{-i \frac{\pi}{4}} \frac{\Gamma(\frac{1}{4}- \frac{1}{4} i \epsilon_\text{R}) }{\Gamma(\frac{1}{4}+ \frac{1}{4} i \epsilon_\text{L}) }
  + \frac{E_\mathrm{R}^\mathrm{sad}}{E_\mathrm{L}^\mathrm{sad}}e^{-i \frac{3\pi}{4}} \frac{\Gamma(\frac{3}{4}- \frac{1}{4} i \epsilon_\text{R}) }{\Gamma(\frac{3}{4}+ \frac{1}{4} i \epsilon_\text{L}) } \Big] 
  \nonumber\\
  & \qquad
  -\hbar\, \text{Im} \frac{\partial}{\partial E} \ln \Big[
  \frac{\Gamma(\frac{1}{4}- \frac{1}{4} i \epsilon_\text{R}) }{\Gamma(\frac{1}{4}- \frac{1}{4} i \epsilon_\text{L}) }
  +\frac{E_\mathrm{R}^\mathrm{sad}}{E_\mathrm{L}^\mathrm{sad}}\frac{\Gamma(\frac{3}{4}- \frac{1}{4} i \epsilon_\text{R}) }{\Gamma(\frac{3}{4} - \frac{1}{4} i \epsilon_\text{L}) }
  \Big] 
  \label{eq:tau-dR-asym}
  .
\end{align}
In obtaining the semi-classical contribution to these expressions, care must be taken to  keep the beamsplitter region fixed, which means ignoring the energy dependency in $\mathcal{X}_{\text{L}}$ and $\mathcal{X}_{\text{R}}$ when differentiating \eq{eq:theta} and \eq{eq:rho}. 
This is because when we consider the delay time, we consider a long wavepacket (energetically narrow) \cite{hauge:tunneling} and the boundary positions of the beamsplitter region are fixed in the energy window of the packet. 

\end{widetext}

Expressions for the phases $\widetilde{\theta}$ and $\widetilde{\rho}$ and the corresponding delay times when electrons impinge from the right are obtained by swapping indices  L $\leftrightarrow$ R in the above.

\subsection{Transmission and reflection probabilities}
In Hamiltonian $H_1$, the probability flux of the scattering state $\psi_{E}$ is
\begin{equation}
  \label{eq:J}
  J = 
  \begin{cases}
    2 E_\mathrm{L}^\mathrm{sad} \text{Im} \, \big( \psi_{E}^*
    \frac{\partial \psi_{E}}{\partial \mathcal{X}} \big)
    & \text{for } \mathcal{X} <0 \\
    2 E_\mathrm{R}^\mathrm{sad} \text{Im} \, \big( \psi_{E}^*
    \frac{\partial \psi_{E}}{\partial \mathcal{X}} \big)
    & \text{for } \mathcal{X} >0 
  \end{cases}
\end{equation}
We therefore obtain the flux for the incoming and outgoing asymptotic forms (input from the left)
\begin{equation}
  \label{eq:J-asymp}
  \begin{aligned}
    |J^{\text{(in)}}(\mathcal{X}<0)|
    &= 2 E_\mathrm{L}^\mathrm{sad} \big(1 + \frac{\epsilon_{\text{L}}}{2 \mathcal{X}^2} \big) |\psi^{\text{(in)}}_{E}(\mathcal{X})|^2    
    \\
    |J^{\text{(out)}}(\mathcal{X}>0)|
    &= 2 E_\mathrm{R}^\mathrm{sad} \big(1 + \frac{\epsilon_{\text{R}}}{2 \mathcal{X}^2} \big) |\psi^{\text{(out)}}_{E}(\mathcal{X})|^2
    \\
    |J^{\text{(out)}}(\mathcal{X}<0)|
    &= 2 E_\mathrm{L}^\mathrm{sad} \big(1 + \frac{\epsilon_{\text{L}}}{2 \mathcal{X}^2} \big) |\psi^{\text{(out)}}_{E}(\mathcal{X})|^2
    \end{aligned}
\end{equation}
Then, using Eqs.~(\ref{eq:t-asymp}) and (\ref{eq:r-asymp}), the transmission and reflection probabilities of the saddle are found to be
\begin{equation}
  \label{eq:T-R}
  \begin{aligned}
    T &= \frac{E_\mathrm{R}^\mathrm{sad}}{E_\mathrm{L}^\mathrm{sad}} e^{-\frac{\pi}{4} (\epsilon_\text{R}-\epsilon_\text{L}) }
     \Big| \frac{ e^{-i \frac{\pi}{4}} \frac{\Gamma(\frac{1}{4}- \frac{1}{4} i \epsilon_\text{R}) }{\Gamma(\frac{1}{4}+ \frac{1}{4} i \epsilon_\text{R}) }
    -e^{-i \frac{3\pi}{4}} \frac{\Gamma(\frac{3}{4}- \frac{1}{4} i \epsilon_\text{R}) }{\Gamma(\frac{3}{4}+ \frac{1}{4} i \epsilon_\text{R}) } }
  { \frac{\Gamma(\frac{1}{4}- \frac{1}{4} i \epsilon_\text{R}) }{\Gamma(\frac{1}{4}- \frac{1}{4} i \epsilon_\text{L}) }
    +\frac{E_\mathrm{R}^\mathrm{sad}}{E_\mathrm{L}^\mathrm{sad}}\frac{\Gamma(\frac{3}{4}- \frac{1}{4} i \epsilon_\text{R}) }{\Gamma(\frac{3}{4} - \frac{1}{4} i \epsilon_\text{L}) } }\Big|^2 \\
  R&=
   \Big| \frac{ e^{-i \frac{\pi}{4}} \frac{\Gamma(\frac{1}{4}- \frac{1}{4} i \epsilon_\text{R}) }{\Gamma(\frac{1}{4}+ \frac{1}{4} i \epsilon_\text{L}) }
    + \frac{E_\mathrm{R}^\mathrm{sad}}{E_\mathrm{L}^\mathrm{sad}} e^{-i \frac{3\pi}{4}} \frac{\Gamma(\frac{3}{4}- \frac{1}{4} i \epsilon_\text{R}) }{\Gamma(\frac{3}{4}+ \frac{1}{4} i \epsilon_\text{L}) } }
  { \frac{\Gamma(\frac{1}{4}- \frac{1}{4} i \epsilon_\text{R}) }{\Gamma(\frac{1}{4}- \frac{1}{4} i \epsilon_\text{L}) }
    +\frac{E_\mathrm{R}^\mathrm{sad}}{E_\mathrm{L}^\mathrm{sad}}\frac{\Gamma(\frac{3}{4}- \frac{1}{4} i \epsilon_\text{R}) }{\Gamma(\frac{3}{4} - \frac{1}{4} i \epsilon_\text{L}) } }\Big|^2
  \end{aligned}
\end{equation}
%

\end{document}